\documentclass[useAMS,usenatbib]{mn2e}

\usepackage{graphics}

\title[MOST photometry of eccentric binary HD~313926]
{Discovery of the strongly eccentric, short-period 
binary nature of the B-type system
HD~313926 by the MOST satellite\thanks{Based on data from MOST, 
a Canadian Space Agency mission jointly operated by 
Dynacon Inc., the University of Toronto Institute for 
Aerospace Studies and the University of 
British Columbia, with the assistance of the University 
of Vienna), and on data from the David 
Dunlap Observatory, University of Toronto.}}

\author[Slavek M. Rucinski et al.]
{Slavek Rucinski$^1$,  Rainer Kuschnig$^2$, Jaymie M. Matthews$^2$,
\newauthor
Wojtek Dimitrov$^3$, Theodor Pribulla$^4$, David B. Guenther$^5$, 
\newauthor
Anthony F. J. Moffat$^6$, Dimitar Sasselov$^7$, Gordon A. H. Walker$^2$, 
\newauthor
Werner W. Weiss$^8$ \\
$^1$David Dunlap Observatory, University of Toronto, P.O.~Box 360, 
Richmond Hill, Ontario, L4C~4Y6, Canada\\
$^2$Department of Physics \& Astronomy, University of 
British Columbia, 6224 Agricultural Road, \\  
Vancouver, B.C., V6T~1Z1, Canada\\
$^3$Astronomical Observatory of  A.\ Mickiewicz University, 
ul.\ S{\l}oneczna 36, 60-286 Pozna\'n, Poland\\
$^4$Astronomical Institute, Slovak Academy of Sciences, 
059~60 Tatransk\'a Lomnica,  Slovakia\\
$^5$Institute for Computational Astrophysics, 
Department of Astronomy and Physics, 
Saint Marys University, \\  Halifax, N.S., B3H~3C3, Canada\\
$^6$Observatoire Astronomique du Mont M\'{e}gantic, 
D\'{e}partment de Physique, Universit\'{e} 
de Montr\'{e}al, C.P.6128, \\  Succursale: Centre-Ville, 
Montr\'{e}al, QC, H3C~3J7, Canada\\
$^7$Harvard-Smithsonian Center for Astrophysics, 
60 Garden Street, Cambridge, MA 02138, USA\\
$^8$Institut f\"{u}r Astronomie, Universit\"{a}t Wien, 
T\"{u}rkenschanzstrasse 17, A-1180 Wien, Austria
}

\date{Accepted --.      Received -- ;      in original form --}

\pubyear{2007}

\begin{document}
\label{firstpage}
\maketitle

\begin{abstract}
The MOST photometric space mission discovered an eclipsing 
binary among its guide stars in June 2006 which 
combines a relatively large eccentricity 
$e = 0.20$ with an orbital period of only 2.27 days. 
HD 313926 appears to consist of two early-type stars 
of spectral type B3 -- B7. 
It has a largest eccentricity among known early-type binaries 
with periods less than 3.5~d. Despite the large components 
indicated by its spectral type and light curve model, 
and its short period, the orbit of HD~313926 has not yet 
circularised so it is probably very young, even compared 
to other young B stars.
\end{abstract}

\begin{keywords}
stars: eclipsing -- stars: binary -- stars: evolution -- photometry: 
spacebased
\end{keywords}

\section{Introduction}

Tidal dissipation is known to lead to circularization of 
close binary star orbits. In hot (radiative
equilibrium) stars, the dominant dissipative mechanism 
is thought to be the dynamical tides. Its efficiency is 
expected to show a very strong dependence on the relative 
size of the stars, with the time scale 
$t_{circ}^R \propto (r/a)^{-21/2}$ ($r$ is the radius
and $a$ is the mean separation of the components), even stronger
than for cool (convective equilibrium) stars,
$t_{circ}^C \propto (r/a)^{-8}$ \citep{z1975,z1977,z2005}.
However, coefficients in the two proportionalities differ by many 
orders of magnitude with the latter efficiency being 
much higher than the former. Therefore, as pointed out
in a review by \citet{z1992}, a low-mass, cool
secondary component can contribute most of the dissipation so
that the range of possibilities here is very wide.

The main tool in studies of orbit circularization
is the period -- eccentricity distribution and its evolution with time.
A whole meeting \citep{DM1992}, jokingly referred to by its editors 
as ``The $e - \log P$ Workshop'', was devoted to this subject.
Twelve years after that meeting, a focussed workshop \citep{CGZ2005} 
discussed apparent limitations of the current theories which
point at a need of additional, more efficient dissipation
mechanisms in addition to those included in the classic theory
of Zahn. 
The discovery of a short-period eccentric binary by the MOST 
space mission adds an observation to an interesting 
region of $e - \log P$ parameter space.

\section{MOST observations}  
\label{sect:disc}

MOST (Microvariability \& Oscillations of STars) is a microsatellite 
housing a 15-cm telescope which feeds a CCD photometer through a single 
custom broadband optical filter (350 - 700 nm). The pre-launch 
characteristics of the mission are described by \citet{WM2003}
and the initial post-launch performance by \citet{M2004}.
MOST is in a Sun-synchronous polar orbit (820 km altitude) 
from which it can monitor some stars for as long as 2 months without 
interruption.  The instrument was designed to obtain highly 
precise photometry of bright stars through Fabry Imaging. 
Since launch, its capabilities have been expanded to obtain photometry 
of the guide stars used to orient the spacecraft, with the same 
time coverage as the Primary Science Target. 

HD~313926 ($V = 10.7$, sp. B9) was one of the guide stars 
in the field of the Wolf-Rayet star WR 111, which was the primary 
target observed by MOST in June 2006. 
CCD readings were obtained in rapid sequence with an exposure time 
of 1.5~s, set by the Attitude Control System (ACS) requirements 
of spacecraft pointing.  For the science measurements, these images 
were stacked on board the satellite, so that the effective 
exposure time was 15 s, at a sampling rate of once every 20 s.
The final MOST photometry 
consists of 4,906 readings binned at intervals of 2 -- 6 minutes,
covering about 23 days of nearly continuous monitoring, with
practically no gaps in time.

\begin{figure}
\begin{center}
\scalebox{0.30}{\includegraphics{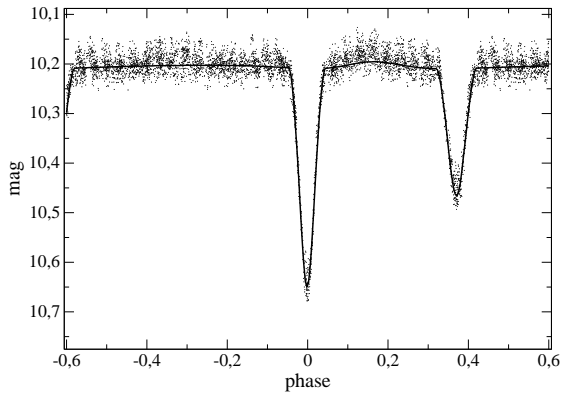}}
\caption{\label{fig-most}  The MOST photometry of HD~313926 in 2-min bins, 
phased to a period of 2.27038~d.  The solid curve is the best 
fit from our light curve synthesis model.} 
\end{center}
\end{figure}

The eclipsing binary nature of HD~313926 was 
obvious early on in the observations. The photometric data were  
phased to the period $P = 2.27038$~d (see ephemeris below), 
as shown in Figure~\ref{fig-most}. The photometry can 
be downloaded from the MOST Public Data Archive on the 
Science page of {\sf www.astro.ubc.ca/MOST}. 

Like many faint MOST guide stars, rather little is available 
in the literature about HD~313926. We summarise the published 
characteristics of the star in Table~\ref{tab-lit}. The Tycho-1 
\citep{hip} parallax is of very low quality and Simbad 
does not give a source for the listed spectral type of B9. 
The star is almost exactly in the Galactic equator, so 
the colour index $B-V = 0.38$ from Tycho-2 \citep{Tycho2} may be due 
to large interstellar reddening. If the star is nearby, without any reddening, 
then the colour corresponds to approximately F3V.  
Therefore, the published spectral type of B9 requires verification. 

\begin{table}
\begin{scriptsize}
\caption{\label{tab-lit} Information from the literature about HD~313926. 
Mean standard errors are given in parentheses (note: Tyc-1 = Tycho-1, 
Tyc-2 = Tycho-2).}
\begin{center}
\begin{tabular}{ll} 
Parameter & Data \\[1.5ex]
Designations & HD~313926\\ 
             & CPD$-21\degr 6659$ \\
             & GSC~06276--01849 \\
Position (J2000) & $\alpha = $18:09:07.93, $\delta = -21$:28:24.8 \\
Galactic coordinates (deg) & $l=9.12$, $b=-0.85$ \\
Photometry (Tyc-1) & $B-V=0.372$ \\
Photometry (Tyc-2) & $V_T=10.602$ (0.062) \\
                   & $B_T=11.059$ (0.062) \\
                   & $B-V=0.388$ \\
Data from Simbad & $V=10.7$, \ \ Sp Type: B9 \\
Parallax (Tyc-1) & $\pi = 90.4\,(55.0)$ \\
Proper motion (Tyc-2) & PM$_{\rm RA} = 0.8\, (2.8)$ mas/yr \\  
              & PM$_{\rm dec} =-1.6\, (2.9)$ mas/yr\\
\end{tabular}
\end{center}
\end{scriptsize}
\end{table}

\section{The binary ephemeris}
\label{time}

It is clear from Figure~\ref{fig-most} that HD~313926 
is an eccentric binary, with the secondary eclipse at 
phase 0.38 relative to the primary eclipse at phase 0.00.

\begin{figure}
\begin{center}
\scalebox{0.45}{\includegraphics{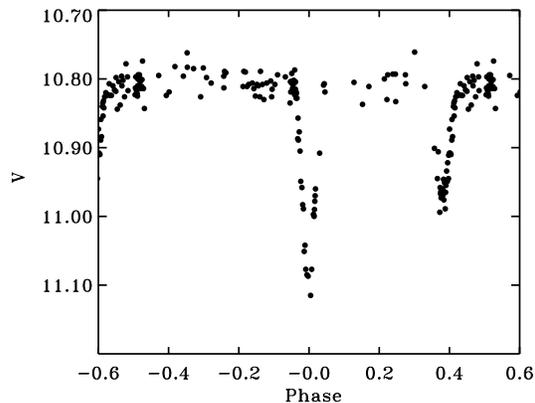}}
\caption{\label{fig-asas} The phased ASAS light 
curve of HD~313926 for the 2003 season (JD 2,452,743 -- 2,452,952).
}
\end{center}
\end{figure}

HD~313926 had been observed photometrically by the 
ASAS photometric survey \citep{Poj05,BP2006}, 
but the large displacement of 
the secondary eclipse went unnoticed in this survey. 
The available $V$-band ASAS observations (482 in number) 
were collected during five consecutive seasons between 
June 2001 and July 2004. Typical data are shown in 
Figure~\ref{fig-asas}, from the 2003 season.  
Note the difference in phase coverage of the ground based 
data spanning 209 days compared to that of the MOST data 
in Figure~\ref{fig-most} collected over only 23 days. 
The depths of the eclipses are different in 
Figures~\ref{fig-most} and \ref{fig-asas}
because of the differences in the filters used 
for the respective photometric measurements.

The ASAS photometry permits determination 
of three seasonal moments of primary eclipses and, 
as a result, an improvement of the primary 
eclipse timing solely from the MOST data
(Table~\ref{tab-asas}). We determine from the 
MOST and ASAS data an ephemeris of:
\[ Min\,I = 2,453,893.0710(15) + 2.27038 (22) \times E \]
where the errors in the last significant digits are given 
in parentheses. There is no clear indication of any 
apsidal motion in the secondary eclipse times from the ASAS 
and MOST data.

\begin{table}
\begin{scriptsize}
\caption{\label{tab-asas}
Eclipse timing for HD~313926 (A = ASAS, M = MOST).}
\begin{center}
\begin{tabular}{cccccc} 
JD pri &   E   & O-C (d)  & JD sec  & Sec ph.  & Source \\[1.5ex]  
2,452,054.066 & $-810$ & $+0.002$ & 2,452,054.920 & 0.377 & A 2001 \\
2,452,362.828 & $-674$ & $-0.007$ & 2,452,363.695 & 0.379 & A 2002 \\
2,452,846.432 & $-461$ & $+0.006$ & 2,452,847.300 & 0.385 & A 2003 \\
2,453,893.071 & 0      &    0     & 2,453,893.912 & 0.370 & M 2006 \\
\end{tabular}
\end{center}
\end{scriptsize}
\end{table}

\section{Light curve synthesis}
\label{sect:synth}

In the MOST measurements, the 
primary eclipse is about 0.45 mag.\ deep and 
the secondary eclipse about 0.25 mag.\ deep.
Both eclipses are relatively short and partial so that 
a light curve synthesis solution is rather 
poorly constrained.  A preliminary solution of the MOST 
light curve has been obtained with the program PHOEBE \citep{PZ2005}, 
which is a convenient interface to the popular Wilson-Devinney 
light curve synthesis code. 

Because the spectroscopic mass ratio is unknown, 
we assumed that the system consists of 
identical-mass components, $q=1$, which is basically the 
``default'' value. There is very little information 
in the light curve 
(except the weak ellipsoidal variability apparent in 
Figure~\ref{fig-most}) to constrain $q$. The equal-mass 
assumption is spectroscopically testable 
(see Section~\ref{sect:spec} below). However, 
most of the parameters in the solution depend only weakly 
on the assumed mass ratio. We note that 
the two stars differ in their surface brightness. 
The two stars have similar dimensions, 
but the primary component (eclipsed at primary minimum) 
appears to be slightly brighter and thus also brighter for
similar radii; therefore, the components cannot be both
simultaneously on the Main Sequence (although our solution
with $q=1$ must be treated with considerable reservation).

The best fit, shown in Figure~\ref{fig-most}, 
with parameters listed in Table~\ref{tab-para}, 
reproduces the observed light curve quite well. 
In particular, the elliptical orbit causes a stronger 
ellipsoidal distortion of the components at the first 
maximum near phase $0.15$. The components 
at that phase are closest to each other and thus most deformed. 

\begin{figure}
\begin{center}
\scalebox{0.5}{\includegraphics{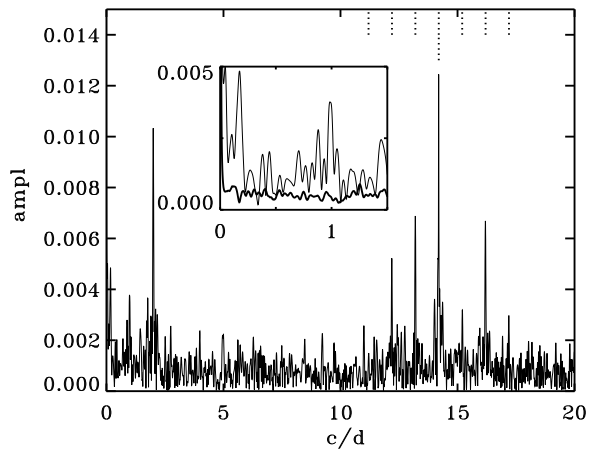}}
\caption{\label{fig-freq} The Fourier-amplitude 
spectrum of residuals of the MOST photometry 
from the theoretical light curve (Figure~\ref{fig-most}).
The spectrum shows only known sky background modulation 
artefacts due to MOST's orbit (14.1994 c/d and 1 cycle/day 
modulation of the scattered Earthshine, marked
by dotted tick marks in the upper part of the figure). 
The inset is an expanded plot of the low-frequency end 
of the spectrum, showing no synthesis model imperfections 
at the binary orbital frequency of 0.44 c/d or its harmonics.
The thick line in the insert gives the error level; 
it is not plotted on the main diagram for clarity, but 
is similar in height throughout the whole spectrum.
}
\end{center}
\end{figure}

To check for a presence of systematic deviations from the
light curve model in Figure~\ref{fig-most}, both in terms
of possible stellar variability and in terms of imperfections of 
the light curve model (which would show at the frequency
0.44 cycles per day), we analyzed the periodic content
of the residuals of the MOST data from the light curve model. 
Figure~\ref{fig-freq} shows the Fourier amplitude spectrum.
The spectrum does not show any variability beyond instrumental
effects due to modulation of scattered Earthshine in the 
MOST instrument focal plane 
as MOST's Sun-synchronous orbit carries 
it above a similar albedo feature on the Earth 
after 1 day. None of these are aliases due to gaps in the data.
The dominant peak occurs at a frequency of 2 cycle/day, 
with another weaker peak at 1 c/d. The other prominent 
peaks occur at the MOST satellite 
orbital frequency of 14.1994 c/d ($P = 101.4$ min) and 
cycle/day side lobes. 

\begin{table}
\begin{scriptsize}
\caption{\label{tab-para}
The preliminary geometric parameters of HD~313926.
The radii $r$ are in units of the mean separation and the
luminosities $L$ are in units of the total flux.}
\begin{center}
\begin{tabular}{lcc} 
Parameter &  Value   & rms error \\[1.5ex]
Mass ratio $q$        &    1.0   &   fixed   \\
Inclination [deg]     &   82.8   &  0.1 \\
Eccentricity          &  0.209   & 0.001 \\
Long.\ periastr. $\omega$ [rads]       & 2.85 &  0.02 \\
$r_1$                 &  0.157   & 0.002 \\
$r_2$                 &  0.160   & 0.002 \\
$L_1$                 &  0.580   & 0.003 \\
$L_2$                 &  0.420   & 0.003 \\
\end{tabular}
\end{center}
\end{scriptsize}
\end{table}

\section{Spectroscopy of HD~313926}
\label{sect:spec}

A few classification-resolution spectra of HD~313926 
were obtained with the 1.88m telescope of the David 
Dunlap Observatory (DDO), at a mean wavelength of 4200~\AA,
covering the region 3900 -- 4650 \AA\ with 
a resolution of about 1.2~\AA. Because the star 
is visible from DDO at an elevation of only 
$25\deg$ above the southern horizon, over the very bright 
Toronto sky, the spectra are of relatively low quality
with the strongly attenuated blue part and with a large noise
due to the bright background subtraction.  
Our classification must be considered very preliminary, 
to be confirmed from a spectrograph in the southern hemisphere.

We have been able to confirm the early type of the 
components from the presence of the He~I lines 
$\lambda4387$ and $\lambda4471$. The range of 
admissible spectral types is B3 to B7, which is even earlier 
than given by Simbad.  Therefore, the binary appears to 
consist of massive, large stars, and the large $B-V$ colour must be
due to the interstellar reddening.

A simple reality check on the size of the components, 
from Kepler's Third Law and the geometrical 
elements in Table~\ref{tab-para}, as well as the main 
sequence mass -- radius relation, suggests 
components with radii of about $4 - 6$ $R_\odot$ and masses 
of $10 - 15$ $M_\odot$.  These ranges are consistent with 
spectral types of B2 to B5. 

A thorough radial velocity (RV) study of HD~313926 
would be useful.  This calls for a 1.5--2m-class telescope in 
the southern hemisphere with a spectrograph of resolving power 
$R \sim 10,000$. Such a study would provide masses 
and absolute dimensions, permitting a full 
characterisation of the system. Based on the MOST light curve, 
the RV semi-amplitudes are expected to be of the order 
$K_1 \simeq K_2 \simeq 250$ km/s for mid-B stars like those 
inferred to be in HD~313926 (${\sim}150$ km/s for mid-F stars). 
However, the overall solution will still 
not be extremely accurate because of the partial eclipses.  

\begin{figure}
\begin{center}
\scalebox{0.45}{\includegraphics{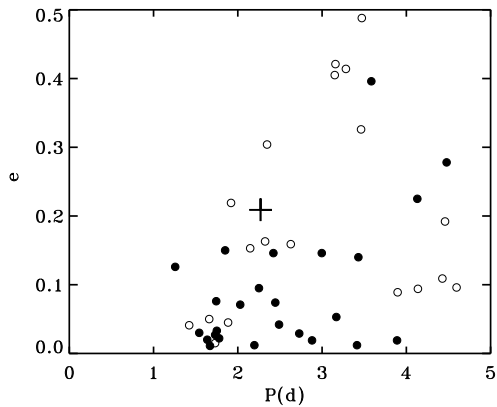}}
\caption{\label{fig-dist}
The distribution of the eccentricity $e$ as a function of the 
orbital period $P$ for close binaries, based on the catalogue 
of \citet{hege2005}. Filled circles are binaries of
 spectral types O to B, while open circles are of types A to G.  
HD~313926 is marked by a cross.}
\end{center}
\end{figure} 

\section{The large orbital eccentricity of HD~313926 in perspective}
\label{sect:ecc}

Close binaries with similarly short periods and large 
eccentricities do exist, but for periods around 
2 days, systems earlier than A0 have 
eccentricities typically below $e \simeq 0.15$. 
In Figure~\ref{fig-dist}, we show the distribution 
of measured values of $e$ vs.\ orbital period $P$ 
for periods less than 5 days based on data from the 
most recent version of the \citet{hege2005} 
catalogue. The figure suggests that the upper envelope 
for early-type stars may be flatter than that 
for spectral types later than A0; for late spectral types
of the same orbital period, 
stars are smaller and the tidal dissipation 
correspondingly weaker. HD~313926, of spectral type 
B3 -- B7, lies near the extreme limit of eccentricity,  
compared to 22 other studied early-type stars with periods
less than 3.5~d (of which 16 have $e < 0.1$). 

In summary, HD~313926 defines a new position of the
upper envelope for eccentricities of early-type stars
at short orbital periods. 
Although all early-type stars are young, this 
pair of stars may be very young compared to the other 
youngsters. Unfortunately, we have no idea what was its 
initial eccentricity, to judge how much the orbit may 
have evolved in a short time. With the current preliminary
geometrical parameters, we also cannot predict how long
the circularization process for HD~313926 would take.
The system definitely requires further
attention in efforts to estimate the circularization
rates in massive, early type stars.

\section*{Acknowledgments}
The Natural Sciences and Engineering Research Council of
Canada supports the research of DBG, JMM, AFJM, and SMR.
Additional support for AFJM comes from FQRNT (Que´bec). 
RK is supported by the Canadian
Space Agency and WWW is supported by the Austrian Space
Agency and the Austrian Science Fund (P17580). Visits of
WD and TP to DDO were partly supported by 
the Polish grant 1~P03D~025~29 of Dr.\ T.\ Kwiatkowski and by 
the Slovak Academy of Sciences VEGA grant 2/7010/7. 
This research has made use of the SIMBAD database,
operated at CDS, Strasbourg, France and NASA's Astrophysics 
Data System Bibliographic Services.

\end{document}